\documentclass[a4paper,twocolumn,fleqn]{article}

\usepackage{ist}
\usepackage{graphicx}
\usepackage{subcaption}
\usepackage{lipsum}
\usepackage{amsmath}
\usepackage{url}
\usepackage{float}
\usepackage[belowskip=4pt,aboveskip=6pt]{caption}
\usepackage{xurl}
\usepackage{hyperref}
\usepackage{xcolor}

\title{Adaptive Color Grading}

\author{Trevor D. Canham \footnotemark, Abhijith Punnappurath $^2$ and Michael S. Brown $^{2}$ \\
        $^1$York University, Canada; $^2$AI Center-Toronto, Samsung Electronics}

\date{}

\begin{document}

\maketitle
\footnotetext{Work done during an internship at Samsung AI Center-Toronto} 
\thispagestyle{empty}


\begin{abstract}
Independent control of tonescale regions (e.g., shadows, highlights) is essential for painters, photographers and cinematographers to bring 2D images to life.
In image manipulation software this is most directly addressed by color grading modules, which use intensity thresholds to segment distinct illumination regions for local manipulation.
In this work we develop an open source color grading tool and use it to annotate a large dataset of video frames with tonescale region thresholds.
Using these thresholds we conduct modeling experiments with strategies based on both practitioners' conventional wisdom and machine learning.
Results show that K-nearest neighbors is an effective prediction strategy, outperforming state-of-the-art end-to-end methods for image enhancement.
This outcome demonstrates the benefit of focusing on a compact set of core parameters when modeling creative stylization processes. Our adaptive color grading interface and data are available \href{https://github.com/SamsungLabs/adaptive-color-grading}{here}.
\end{abstract}

\section{Introduction}
\label{sec:intro}

\begin{figure}[]
\includegraphics[width=1\columnwidth]{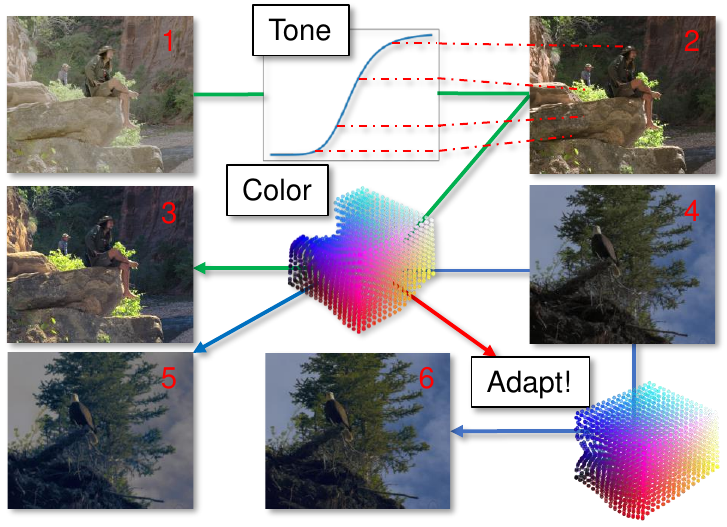}
   \caption{A relatively ``flat'' image appears more voluminous when the contrast between its tonescale regions is increased\textcolor{red}{\textsuperscript{1,2}}. This can be accomplished through tonemapping and chromatic shifts\textcolor{red}{\textsuperscript{2,3}}. However, a static 3D color lookup table tuned to achieve this effect for one image may not work for another\textcolor{red}{\textsuperscript{4,5}}. We present a method for predicting tonescale region thresholds such that these color effects can be adapted to new images\textcolor{red}{\textsuperscript{4,6}}.}
   \label{fig:teaser}
   \vspace{-0.6cm}
\end{figure}

Vocabulary for distinct segments of the tonescale, or the continuous range of light intensity from black to white, has long been part of the lexicon of painters, photographers and cinematographers.
By creating clear distinctions in intensity or color between black, shaded, directly illuminated, and specular highlights, artists can promote a greater sense of volume and contrast in images.
We demonstrate this effect in Figure~\ref{fig:teaser}, where an image first shown in a relatively ``flat'' state appears to increase in volume after manual tone mapping.
The adjacent examples extend the effect with chromatic shifts applied to intensity threshold masks.
These operations constitute the core of the cinema post-production process known as color grading and are also commonly used when stylizing photos and 3D graphics.

Ansel Adams articulated an 11 zone system in terms of relative stops, detailing common scene elements and their textural characteristics as a guide for communicating recommendations on metering, exposure, development and printing photographic film \cite{adams1948}.
Real scenes, however, often break these definitions.
Figure~\ref{fig:teaser} demonstrates one such case, where the color grading of the original image (middle left) is statically applied to a different image (bottom left).
The grading was tuned so that shadows in the first image receive a blue cast and directly illuminated regions a yellow one.
In the second image, however, the 
the intended chromatic separation collapses.
A color grading algorithm adaptive to new sources would therefore be desirable.
While many approaches have been proposed to learn color manipulation styles \cite{bychkovsky11, isola2017image, conde24, serrano24}, their effectiveness at predicting multiple complex, highly subjective editing decisions end-to-end remains unproven.

In this work we demonstrate that by focusing on the compact parameter set of tonescale region thresholds (TRTs), adaptive color grading can be learned and evaluated more effectively and interpretably.
To this end we develop an open source color grading tool and use it to stylize a large-scale dataset of motion picture frames~\cite{hdrvs2026}.
We then conduct extensive experiments with various learning-based approaches.
The resulting model adapts a simple color grading algorithm to new images more effectively than end-to-end methods.
The remainder of this paper reviews related work, presents the above contributions 
and benchmarks these strategies against end-to-end methods.

\section{Related Work}

When the diverse scenes and dynamic illumination conditions of the natural world meet the limited representation capacity of biological and man-made vision systems, the relationship between scene light and visual output is modulated via adaptation.
Harmony with biological adaptation is desirable for man-made vision systems tasked with reproducing human visual experiences (e.g., photography and cinematography).
This challenge has most notably been addressed in camera pipelines in the form of automatic exposure \cite{onzon21, sampat99, yang08}, white balance \cite{afifi19, funt96, liu95} and tone mapping \cite{cyriac21, kuang07, tumblin93}.

More recent works have expanded adaptive image processing to account for full-suite image editing operations including tone mapping, exposure, global color adjustments, local object masks and filters.
The foundational work for this area comes from Bychkovsky et al.~\cite{bychkovsky11}, who arranged a dataset of 5,000 manually retouched images.
They then conducted a series of experiments to predict unseen editing decisions using a broad range of image features as input to an editing style regression model.
Later, more general solutions for learning image-to-image translation were introduced.
One such method is the U-Net \cite{isola2017image}, where multiscale convolutional transforms are learned to match input/output pairs.
Other works aim to produce image-adaptive global color stylization by training two-stage networks that first transform images into context-dependent embeddings, then feed these embeddings into a learned enhancement mechanism (e.g., 3D-LUTs \cite{zeng20} or bilateral grids \cite{kim24}) trained on enhanced references.
A recent work closely related to the present application takes a constrained approach based on global color segmentation, learning independent non-linear functions for separate color name categories \cite{serrano24}.
Another related work uses a multilayer perceptron (MLP) to compress and blend between multiple 3D-LUTs \cite{conde24}.
Common to these works is the goal of learning many editing operations simultaneously from input/output pairs (end-to-end), so none evaluate whether any individual operation is being modeled correctly.

In the cinema industry, color grading is practiced by a small number of skilled workers known as colorists.
These artists use a variety of tools to correct discontinuities caused by inconsistent lighting conditions or shooting errors, highlight important scene elements and beautify movies.
While the origins of this process lie in chemical film development, the pre-modern color grading interface was introduced as a component of the Telecine, a dedicated machine for transferring print film to electronic signals for television broadcast.
The resulting signals could be manipulated by additive (lift), exponential (gamma), and multiplicative (gain) operations, which roughly control dark, middle, and bright tonescale regions respectively.

Recent advances in high dynamic range camera and display technology, however, have extended the number of perceptually distinct tonescale regions achievable in images and rendered this type of legacy control ``too clunky'' for some cinematographers \cite{yedlin16}.
As a result, tools with more fine-grained control regions have been introduced, generally featuring some number of predefined or manually adjustable tonescale regions.
Such controls also appear in photo editing tools like Adobe Lightroom \cite{lightroom}.
While commercial color grading tools are constantly in development, little research is published on how they are developed and used.
A recent work from Tschannerl \& Siragusano~\cite{tschannerl26} aims to automate the shot matching step of color grading by semantically segmenting common scene elements between shots and matching them in terms of exposure, color balance, and black level.
Their model conforms input footage in preparation for later creative adjustments, whereas the present work adapts the creative adjustments themselves to new images.

A more closely related work from Bonneel et al.~\cite{bonneel13} presents a video style transfer method based on the color grading process.
Their transfer mechanism learns a non-linear function for the achromatic channel and chromatic affine transforms applied to three separate tonescale regions.
The tonescale regions are defined as three bands containing an equal number of pixels with cubic falloff and 10\% overlap.
While this solution allows tonescale-region-specific chroma shifts to be defined for a reference style and transferred to a new source, the definition is not informed by any real color grading data.
We address this gap by modeling tonescale region definitions on a large dataset of motion picture frames.
By introducing a model for this compact parameter set, we enable a simple color grading algorithm to adapt to unseen sources.

\section{Methods}

Our methodology to develop an adaptive color grading tool involves three major steps.
First, we developed an open source color grading tool.
Next, we used the tool to define tonescale region thresholds (TRTs) on a large number of still frames from an HDR video dataset \cite{hdrvs2026}, enabling independent chroma offsets across illumination regions.
Finally, we developed and tested a variety of modeling strategies for learning these settings.

\subsection{Color Grading Tool}

\begin{figure}[]
  \includegraphics[width=1\columnwidth]{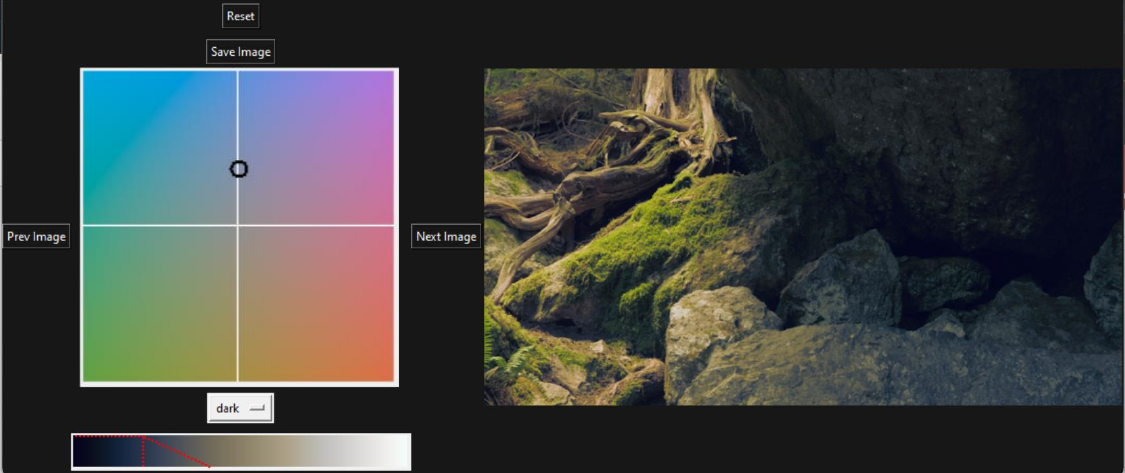}
  \caption{Color grading interface featuring chroma offset, TRT controls and grayscale visualization.}
  \label{Figure:graderUI}
  \vspace{-0.5cm}
\end{figure}

To investigate and model how TRTs are set in practice, we developed an open source color grading tool whose interface is shown in Figure~\ref{Figure:graderUI}.
The interface consists of two primary components: a CIELAB \cite{CIELAB} chroma offset control and a TRT control.
The chroma offset interface shows a limited range of $a^*b^*$ values at an $L^*$ value of 60.
The TRT control interface features an achromatic gradient allowing users to directly visualize chromatic shifts as a function of achromatic intensity.
In the proposed tool, the tonescale is broken into two sets of two overlapping regions defined by $mean(R,G,B)$ intensity.
The first set covers the dark and darkest segments of the tonescale; the second covers the light and lightest segments.
Support functions for the dark $\omega_{d}$ and light $\omega_{l}$ supersets are given in Eqs.~\ref{eq:one} and \ref{eq:two} as functions of intensity $x$, TRTs $\tau_{d}$ and $\tau_{l}$, and falloff slope $m$.
For the dark and darkest regions, $m$ is set to $-5$ and $-10$ respectively; the light and lightest regions are set to $5$ and $10$.
Resulting $\omega$ values are clipped to the $[0,1]$ range as follows:

\begin{equation}
\label{eq:one}
\omega_{d}(x) =
\left\{
    \begin{array}{lr}
        1, & \text{if } x < \tau_{d}\\
        m x + (1-\tau_{d}m), & \text{if } x \geq \tau_{d}
    \end{array}
\right\}
\end{equation}

\begin{equation}
\label{eq:two}
\omega_{l}(x) =
\left\{
    \begin{array}{lr}
        1, & \text{if } x > \tau_{l}\\
        m x + (1-\tau_{l}m), & \text{if } x \leq \tau_{l}
    \end{array}
\right\}.
\end{equation}

\begin{figure}[]
  \includegraphics[width=1\columnwidth]{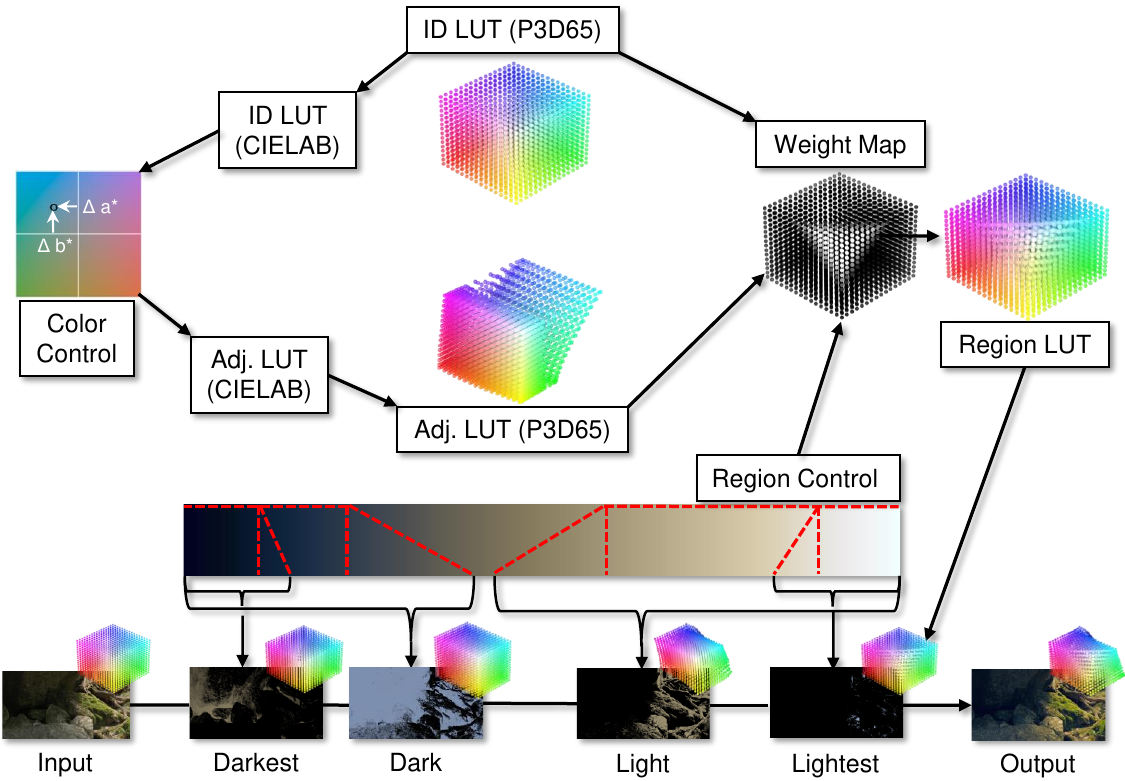}
  \caption{Color grading architecture. Regional LUTs are applied in a nested fashion to an identity LUT which is then applied to return the output image.}
  \label{Figure:graderArch}
\end{figure}

The support functions take the $mean(R,G,B)$ of an identity (ID) 3D-LUT as input and return a single-channel weight map corresponding to each node.
The tool applies user-defined additive $a^*b^*$ channel offsets to four separate ID 3D-LUTs, each transformed from the source RGB encoding to CIELAB.
A weighted average is then taken between the ID 3D-LUT and the chroma-adjusted lookup tables according to the weight maps.
Finally, the 3D-LUTs are applied in a nested fashion to an ID 3D-LUT and the result is applied to the input in a single step.
The complete process is demonstrated graphically in Figure~\ref{Figure:graderArch}. In all of the following experiments we apply 3D-LUTs with $17 \times 17 \times 17$ nodes using trilinear interpolation.

\subsection{Color Grading Dataset}

\begin{figure}[]
  \includegraphics[width=1\columnwidth]{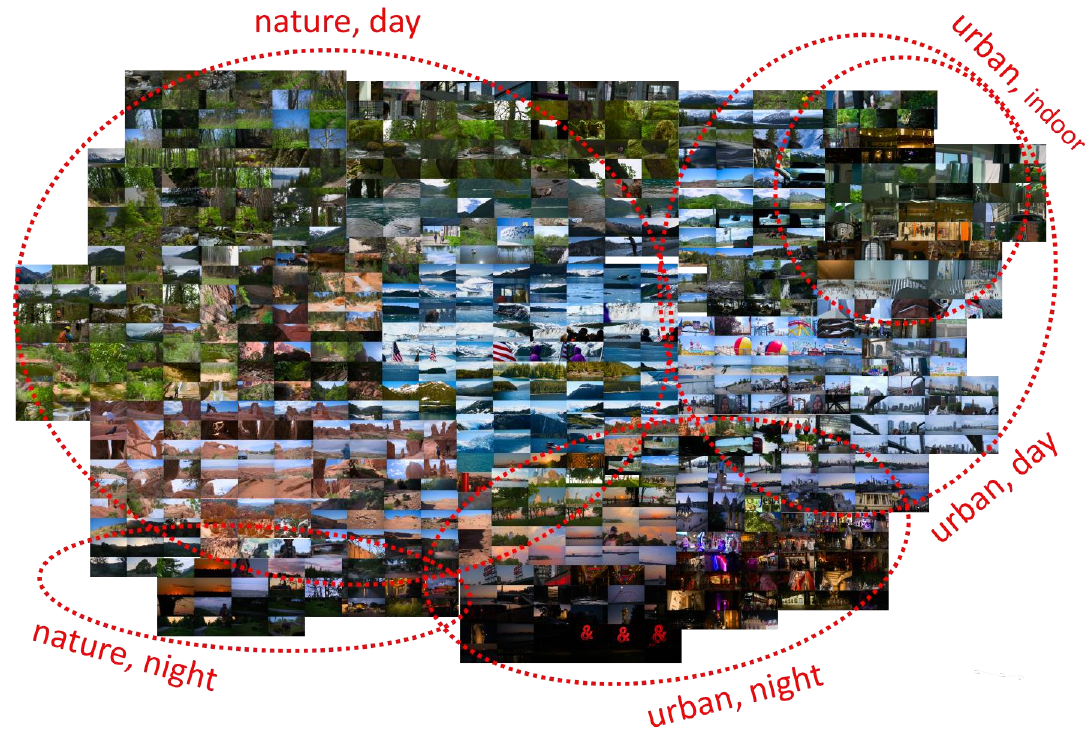}
  \caption{The HDR videographic survey dataset arranged according to scene type.}
  \label{Figure:hvs}
\end{figure}

We manually color graded a large-scale dataset using the tool described above.
We selected 782 individual shots from the High Dynamic Range Videographic Survey (HVS)~\cite{hdrvs2026}
spanning 44 scenes across six distinct geographic regions to cover a variety of illumination conditions and enable analysis of training and testing on individual categories (Table~\ref{tab:subsets}).
Two still frames were extracted from each shot, resulting in a total of 1,564 individual DCI $2K$ images shown in Figure \ref{Figure:hvs}.
Frames were extracted from the RAW video files in a scene-linear representation and converted to a P3D65 \cite{st2113}, 2.4 gamma representation using the luminance-preserving tone mapping of Resolve \cite{resolve}.

\begin{figure}[]
\vspace{-0.3cm}
\centering
  \includegraphics[width=1\columnwidth]{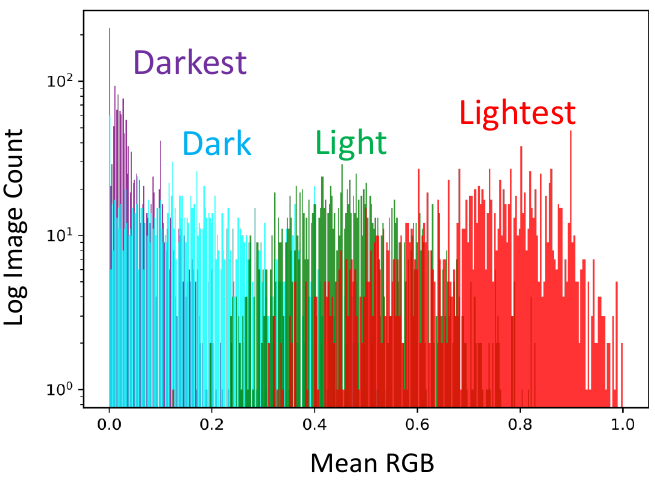}
  \caption{Histogram of tonescale region settings.}
  \label{Figure:dist}
\end{figure}

\begin{table}[]
\vspace{-0.5cm}
\centering
\caption{Breakdown of HVS locations and scene types which differ from the majority day, outdoor, nature scenes.}
\label{tab:subsets}
\begin{tabular}{ |c|c|c| } \hline
 Location subset & Count & Scene type subsets \\ \hline
 Toronto & 90 & Urban \\ \hline
 Utah & 458 & Sunset, Night \\ \hline
 Alaska & 378 & Sunset, Night \\ \hline
 Seattle & 286 & Sunset, Night, Urban, Indoor \\ \hline
 London & 146 & Sunset, Night, Urban, Indoor \\ \hline
 NYC & 206 & Sunset, Urban, Indoor \\ \hline
\end{tabular}
\vspace{-0.4cm}
\end{table}

The intention of color grading was to introduce chroma shifts to directly illuminated and shadowed regions with the goal of increasing chromatic contrast between depth planes.
In other words, the dark and light TRTs were set so that shadows (also known as backlit or indirectly illuminated regions) were separated from directly illuminated objects.
Then, the darkest and lightest TRTs were used to protect the blackest regions, visible illumination sources, specular highlights, and the sky from the chroma shifts.
This was accomplished by applying inverse chroma shifts relative to those applied to the overlapping dark and light tonescale regions.
While some chroma offsets may be preferable to others (e.g., promoting color harmony or accentuating hues present in a particular tonescale region), we observed during annotation that a range of chroma offsets can be appropriate depending on the user's stylistic intentions. 

While this effect was easy to achieve in many images, others required subjective compromises.
In ideal cases, where directly illuminated and shadowed regions are clearly separated in intensity, a range of acceptable threshold settings produces nearly identical graded results.
Many images, however, featured reflectance ambiguities where highly reflective objects under indirect illumination registered with higher intensity than less reflective ones under direct illumination.
Another source of ambiguity was atmospheric conditions like haze, where the shadows on a distant object registered with a higher intensity than directly illuminated objects in the foreground.
Also, certain images featured smooth, extensive penumbras (partially shaded regions) where the boundary between direct and indirect illumination was very gradual, requiring the TRT to be set so its falloff aligned with that of the penumbra.
Finally, images had varying numbers of distinct illumination regions, with a number of cases occupying only two or three of the four defined in the user interface. 
These observations demonstrate why predicting creative stylization is an ill-posed problem.

\subsection{Adaptive Color Grading Model}

\begin{figure}[]
\centering
  \includegraphics[width=1\columnwidth]{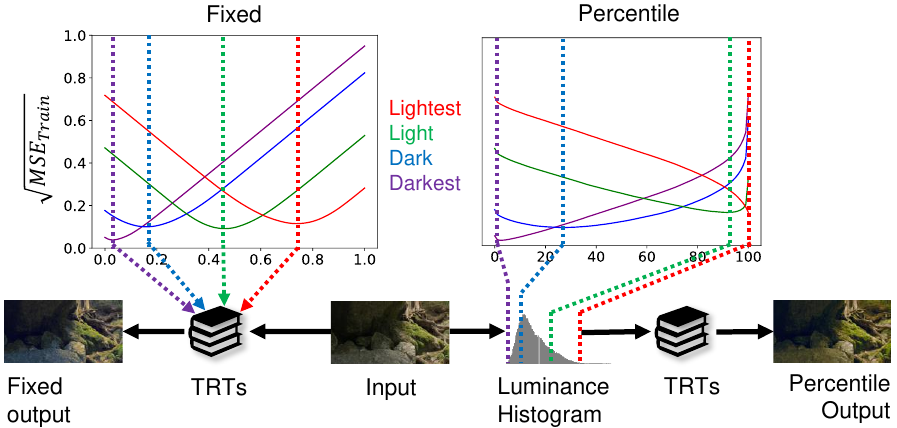}
  \caption{1-D Optimization landscapes for Fixed and Percentile based strategies.}
  \label{Figure:conventional}
  \vspace{-0.5cm}
\end{figure}

We explored several strategies for modeling the TRTs.
First, we tested the conventional wisdom implemented in commercial color grading tools and prior work \cite{bonneel13}.
These materials suggest that properly exposed images on a calibrated camera system can be separated into shadow, midtone and highlight tonescale regions with fixed intensity thresholds or percentiles \cite{vanHurkman13}.
To the authors' knowledge, these claims have never been empirically tested at this scale despite their wide acceptance and propagation.
We defined sets of fixed TRT and percentile values and computed their mean squared error versus the ground truth: directly in the fixed case, and via the luminance histogram of each training image in the percentile case.
Then, the ideal fixed or percentile setting for a given tonescale region was defined as the grid setting with the lowest error.
The ideal fixed settings were used directly in the following experiments, while the ideal percentiles were used to derive image-wise TRTs via the luminance histogram of each test image.
These procedures are outlined in Figure~\ref{Figure:conventional}, which shows their optimization landscapes over the grid search.
The fixed strategy can be conceptualized as an oracle static 3D-LUT, and the percentile strategy as an adaptation of Bonneel et al.'s model \cite{bonneel13} to our data.

\begin{figure}[]
\centering
  \includegraphics[width=1\columnwidth]{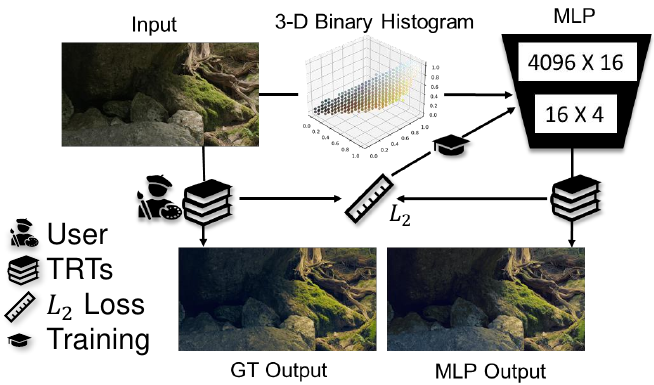}
  \caption{Multi-layer perceptron modeling overview.}
  \label{Figure:mlp}
  \vspace{-0.8cm}
\end{figure}

We next tested two learning-based strategies.
The first was a small multilayer perceptron (MLP) for modeling the TRTs.
An overview of the two-layer network architecture is shown in Figure~\ref{Figure:mlp}.
Ablation experiments (against 12-bit $3 \times 1$ RGB, luminance histograms, and 3D histograms) revealed the best input feature to be a 16-bin binary 3D histogram.
Further ablations indicated an ideal hidden layer size of 16 neurons.
The output layer returns four values corresponding to the TRTs.
Outputs were passed through a sigmoid function to constrain results to the $[0,1]$ range.
Loss was computed as the $L2$ distance between the ground truth and predicted TRTs.
The network was trained with an Adam optimizer \cite{kingma14} for 2,000 epochs at a learning rate of $10^{-6}$, taking around 10 minutes on an Intel i7-1360P CPU.

\begin{figure}[]
\centering
  \includegraphics[width=1\columnwidth]{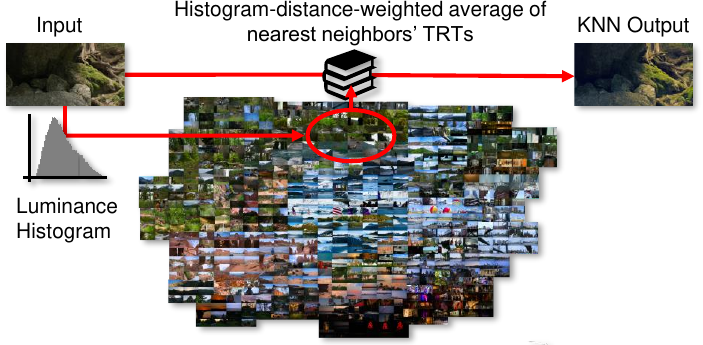}
  \caption{K-nearest neighbors modeling overview.}
  \label{Figure:knn}
  \vspace{-0.8cm}
\end{figure}

Inspired by a similar illumination-based adaptive color manipulation approach for white balance correction \cite{afifi19}, we also modeled the TRTs using K-nearest neighbors regression (Figure~\ref{Figure:knn}).
Ablation experiments revealed best results using 12-bit luminance histograms scaled to unit variance.
At inference time, the predicted TRTs are computed as a weighted average over the 16 training-set neighbors closest to the test histogram by Euclidean distance.
The regression function can be fit and queried in roughly 0.01 and 0.1 s, (respectively, averaged over the various training subsets)  
on an Intel i7-1360P CPU.

\section{Experiments}

We conducted a benchmarking experiment for adaptive color grading models using the dataset annotations described in the previous sections, comparing application-specific modeling strategies against end-to-end methods based on deep learning.
Since the application-specific models predict only TRTs, chroma offsets were fixed for all images at the following $a^*b^*$ levels: Darkest $\{0,-10\}$, Dark $\{0,10\}$, Light $\{0,-10\}$, Lightest $\{0,10\}$, producing roughly blue shadows, yellow directly illuminated regions, and neutral blacks and illumination sources after passing through the color grading process of Figure~\ref{Figure:graderArch}.
This levels the playing field between the application-specific and end-to-end models: while the latter are at a disadvantage for not having access to the precise color grading functions of the open source tool (e.g., the TRT falloff functions of Eqs.~\ref{eq:one} and \ref{eq:two}), TRTs are the only dynamic parameter in the ground truth graded images.
In both experiments, errors are computed between ground truth and predicted graded images in terms of PSNR and $\Delta E_{00}$ \cite{sharma2005}; for application-specific experiments, predicted images are the grader output with predicted TRTs and fixed chroma offsets, while end-to-end experiments use direct model output.

\subsection{Application-specific modeling}

\begin{figure*}[]
    \vspace{-0.2cm}
    \hspace*{0.57cm}
    \begin{subfigure}{\columnwidth}
        \includegraphics[width=2\textwidth]{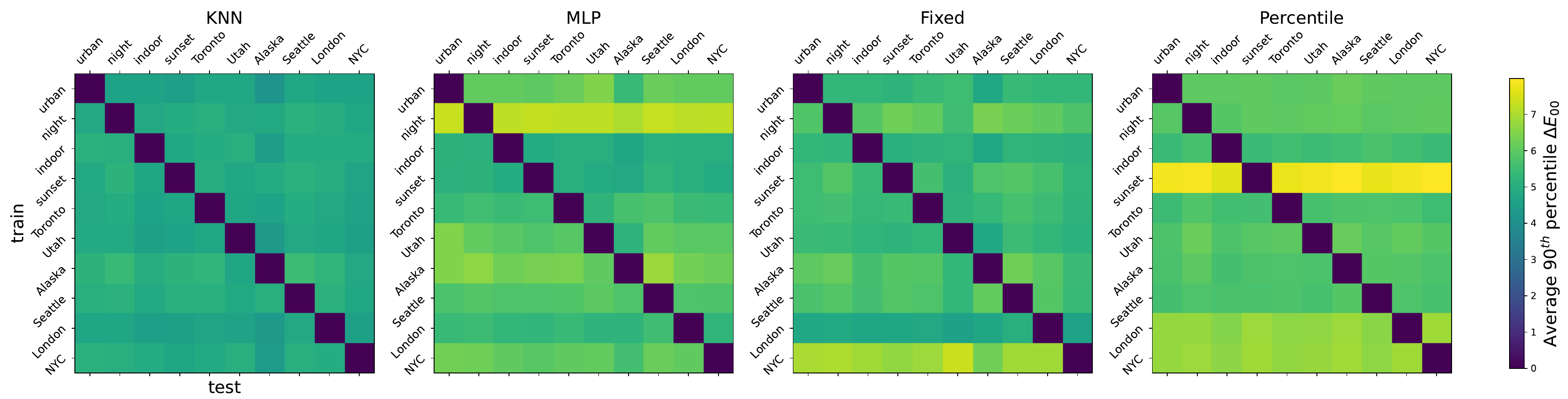}
    \end{subfigure}

    \hspace*{0.56cm}
    \begin{subfigure}{\columnwidth}
        \includegraphics[width=1.88\textwidth]{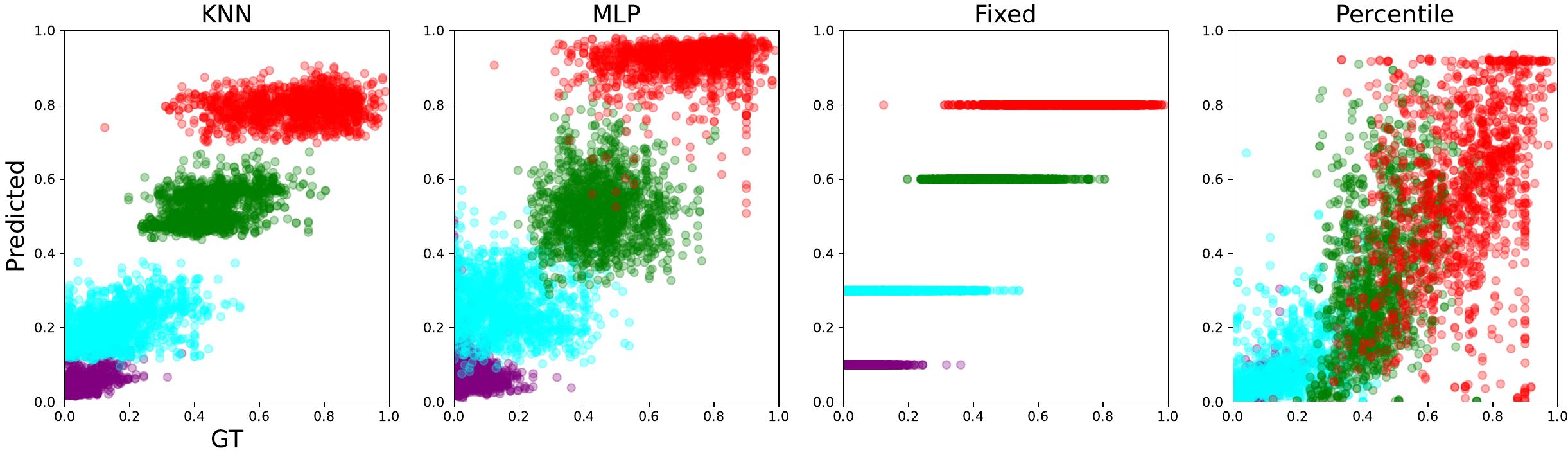}
    \end{subfigure}

    \caption{Top row: 90th percentile $\Delta E_{00}$ averaged across the test set (x-axis). 
    Bottom row: Correlation plots with ground truth darkest (purple) dark (blue) light (green) lightest (red) TRTs plotted against predicted ones by KNN, MLP, Fixed and Percentile strategies.}
    \label{fig:quantitative}
    \vspace{-0.18cm}
\end{figure*}

The rapid training times of the application-specific modeling strategies (Fixed, Percentile, MLP, and KNN) enabled extensive experimentation across all dataset subsets outlined in Table~\ref{tab:subsets}.
Results are shown in Figure~\ref{fig:quantitative}, with a heat-map matrix for all train and test subsets and a correlation plot against ground truth TRTs (training on the NYC subset and testing on the remainder).
Quantitative results are also shown in Table~\ref{tab:quant}, and qualitative results in Figure~\ref{Figure:qualitative}.

Starting with the results matrices for all subsets, particular training sets disproportionately cause failures across all test sets for the MLP, Fixed, and Percentile strategies, while the KNN strategy maintains relatively consistent performance across training sets.
Here, 90\textsuperscript{th} percentile $\Delta E_{00}$ results are shown, indicating the threshold below which 90\% of test image pixels fall, averaged over the full test set.
Due to the small number of predicted parameters (4 TRTs), an interpretable analysis is possible for the application-specific strategies, as shown in the correlation plots of Figure~\ref{fig:quantitative}.
In these plots, each point's x-axis position corresponds to the ground truth setting and its y-axis position to the predicted setting.
While a perfect model would produce a straight diagonal line, the shape of each plot reveals characteristics of its modeling strategy.

The Fixed strategy is the easiest to interpret: ground truth settings spread across the x-axis while predictions remain pinned to fixed points on the y-axis, forming horizontal bands.
The Percentile strategy's correlation plot reveals that while fixed percentile values provide histogram-adaptive behavior, their direct dependence on intensity distribution produces a wide range of erroneous TRTs.
The MLP results mostly mirror the variance of the ground truth TRTs, forming nearly circular clusters indicating that simple neural networks have limited ability to learn the relationship between input image histograms and TRT settings.
In contrast, KNN correlation plots show clearly separated TRT clusters.
While this pattern is not strictly reflected in the ground truth, it is desirable for the user experience, as it ensures region settings remain well-separated.

Quantitatively, the MLP and Percentile strategies underperform the Fixed idealized static LUT (Table~\ref{tab:quant}), while KNN is the only strategy that consistently and significantly surpasses it.
The qualitative results in Figure~\ref{Figure:qualitative} confirm this: KNN-predicted TRTs produce graded images where directly and indirectly illuminated regions are colored as in the ground truth, while MLP, Fixed, and Percentile strategies over-predict the extent of particular tonescale regions and result in graded images of mostly uniform color cast.

\subsection{End-to-end modeling}

The proposed application-specific TRT modeling strategies are also compared against various end-to-end image enhancement paradigms: U-Net \cite{isola2017image}, NILUT \cite{conde24}, and NamedCurves \cite{serrano24}.
Each of these models was trained and tested on dataset subsets identified as relevant from the application-specific experiments.
Specifically, we report results in two configurations: training on the NYC and Night subsets and testing on the remainder, and the inverse (training on the remainder and testing on NYC and Night).

The U-Net baseline \cite{isola2017image} used the original architecture with 32 filters in the first convolutional layer and was trained patch-wise for 100 epochs with a batch size of 32 and a learning rate of 0.01.
The network was trained to predict a residual correction added to the input image to produce the final output.
Training on the NYC subset (206 images) took roughly 85 minutes on an NVIDIA V100 GPU with 16 GB of VRAM.
The NamedCurves method was trained with its default parameters for 100 epochs, taking roughly 56 minutes on an NVIDIA RTX 6000 ADA GPU with 48 GB of VRAM.

\begin{figure*}[]
    \centering
    \includegraphics[width=1.90\columnwidth]{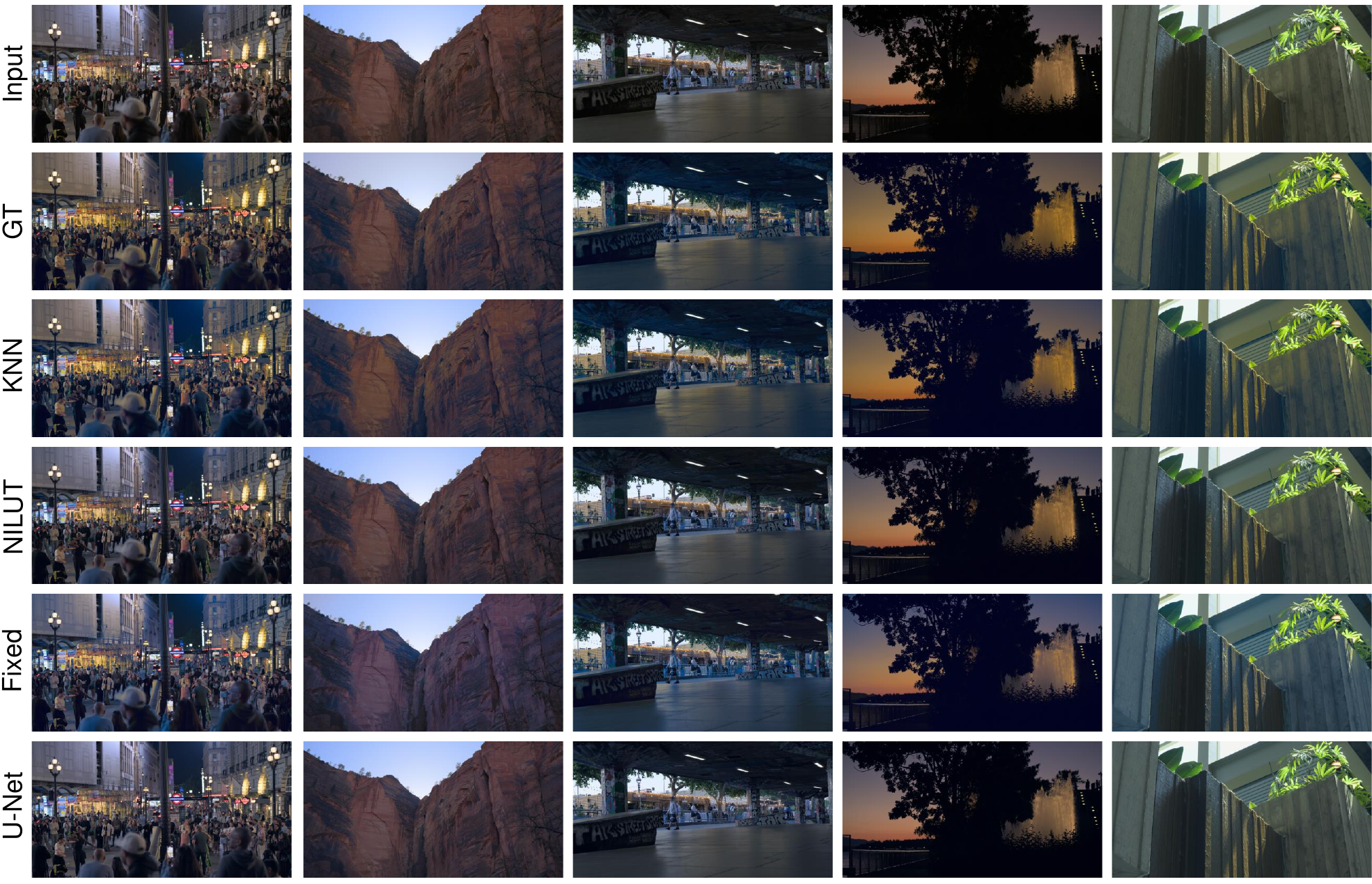}
  \caption{Qualitative results from KNN, NILUT~\cite{conde24}, Fixed and U-Net~\cite{isola2017image} strategies trained on ``NYC'' dataset subset. KNN follows yellow-blue shift regions of GT while other solutions struggle.}
  \label{Figure:qualitative}
\end{figure*}

The NILUT method \cite{conde24} uses an MLP to implicitly parameterize a continuous RGB-to-RGB color transform by mapping input pixel triplets directly to enhanced output values.
We used the MLP architecture and training hyperparameters recommended by the authors.
Unlike the original work, which trains on RGB pairs sampled from LUTs (e.g., via HALD images), we trained directly on real input/target image pairs, taking roughly 100 minutes on an NVIDIA V100 GPU with 16 GB of VRAM.
Qualitative and quantitative results for these methods are shown in Figure~\ref{Figure:qualitative} and Table~\ref{tab:quant} respectively.
Of these three approaches, only NILUT consistently surpassed the fixed static LUT in quantitative measures (Table~\ref{tab:quant}); however, the qualitative results in Figure~\ref{Figure:qualitative} suggest it does so by staying close to the input image to avoid errors.

\begin{table*}[]
\vspace{-1.1cm}
\centering
\caption{Quantitative results of modeling approaches versus end-to-end deep learning approaches. In these experiments models are tested on a subset and trained on the remainder or vice-versa.}
\label{tab:quant}
\begin{tabular}{|c|c|c|c|c|c|c|c|c|} \hline
Train/test & \multicolumn{2}{c}{remainder/NYC} & \multicolumn{2}{|c|}{NYC/remainder} & \multicolumn{2}{|c|}{remainder/Night} & \multicolumn{2}{|c|}{Night/remainder} \\ \hline
 & $\Delta E_{00}$ $\downarrow$ & PSNR $\uparrow$ & $\Delta E_{00}$ $\downarrow$& PSNR $\uparrow$& $\Delta E_{00}$ $\downarrow$& PSNR $\uparrow$& $\Delta E_{00}$ $\downarrow$& PSNR $\uparrow$ \\ \hline
U-Net \cite{isola2017image} & 2.96 & 34.90 & 3.83 & 33.15 & 2.99 & 36.22 & 2.81 & 36.19 \\ \hline
NamedCurves \cite{serrano24} & 3.44 & 32.86 & 4.02 & 32.44 & 4.85 & 30.99 & 3.83 & 32.68 \\ \hline
NILUT \cite{conde24} & 2.65 & 35.46 & 2.86 & 35.80 & 1.62 & 40.84 & 2.56 & 36.87 \\ \hline
Fixed & 3.25 & 34.16 & 3.56 & 34.39 & 1.81 & 40.47 & 2.96 & 35.90 \\ \hline
Percentile \cite{bonneel13} & 3.43 & 33.54 & 3.87 & 34.17 & 2.93 & 37.60 & 3.20 & 34.93 \\ \hline
MLP  & 3.41 & 33.71 & 3.15 & 35.24 & 2.10 & 38.13 & 3.68 & 33.00 \\ \hline
KNN & \textbf{2.40} & \textbf{36.77} & \textbf{2.55} & \textbf{37.53} & \textbf{1.47} & \textbf{42.94} & \textbf{2.50} & \textbf{37.26} \\ \hline
\end{tabular}
\end{table*}

\section{Discussion}

As outlined in the Methods section, predicting tonescale region thresholds (TRTs) requires not only finding signatures of independent illumination regions within image intensity distributions, but also learning the subjective decision-making process of individual users.
This is due to the inherent ambiguity of masking semantic regions with global color segmentation.
That said, this process is computationally amenable and directly applicable to video, as it obviates the need for viewpoint- and scale-dependent object tracking \cite{yang08}.

In the application-specific modeling experiment, we show that various strategies can reliably model TRTs and enable adaptive color grading better than approaches based on conventional wisdom.
On the other hand, end-to-end image enhancement models often failed to surpass the performance of the fixed 3D-LUT.
While their intended application to generically model all photo-finishing operations end-to-end from source/edited pairs is more challenging than the application-specific case, their failure to surpass this static average solution calls into question the prevailing assumption that end-to-end learning is the right approach for highly subjective creative stylization tasks.
Additionally, the relatively long training times and model sizes are prohibitive for personalized on-device applications.
Among the end-to-end approaches, only NILUT was able to surpass static performance, but the qualitative results suggest it does so by staying very close to the input image to avoid errors.

To the authors' knowledge, this work is the first to empirically test the conventional assumptions of tonescale region segmentation, that fixed thresholds or intensity histogram percentiles can roughly approximate illumination regions.
While the fixed strategy performed relatively well in these experiments, we show in Figure~\ref{Figure:dist} that the assumption is too restrictive as thresholds vary significantly in practice.
The percentile strategy, on the other hand, performed the worst of all the application-specific approaches, demonstrating that directly adapting to intensity distributions is problematic for adaptive image processing techniques.
That said, this method was consistent for the darkest threshold, which aptly lines up with the darkest portions of the image.
Other TRTs may require some form of input regularization to reduce dependence on the specific pixel-area proportions of different image intensities (e.g., a shadow is still a shadow whether it occupies 10\% or 90\% of an image).

Another interesting finding was the models' dependence on training data subsets.
In particular, the results matrices of Figure~\ref{fig:quantitative} show that training on the ``Night'' subset caused relatively high test errors for the MLP, the ``NYC'' subset was problematic for the Fixed strategy, and the ``Sunset'' subset was challenging for the Percentile strategy.
In terms of test sets, Table~\ref{tab:quant} shows that testing on ``Night'' images was favorable for quantitative evaluation of most methods aside from NamedCurves \cite{serrano24}, implying that its context extraction block may make it particularly sensitive to out-of-distribution images.

Two limitations bound these conclusions: annotations come from a single user pursuing one specific creative intent 
and TRTs represent only one parameter set within the broader grading workflow. 
Even within these bounds, however, the findings support the central argument of this work: focusing on a compact parameter set yields more effective and interpretable results than learning the full editing process end-to-end.
Furthermore, the lightweight training requirements of strategies like KNN 
using isolated, core parameters makes on-device personalization tractable for complex creative workflows.
Finally, these core parameters have potential as robust constraints for end-to-end approaches.

\section{Conclusion}

A number of image processing algorithms have been proposed to adapt the behavior of cameras and post-processing pipelines to input image characteristics.
In this work, we introduced an adaptive mechanism for color grading.
This was accomplished by developing an open source color grading tool, annotating a large dataset of images, and learning to predict tonescale region thresholds.
In doing so, we experimented with a series of application-specific modeling techniques drawing on both machine learning and conventional wisdom from photographic and cinematographic practices, upon which existing commercial color grading tools are based.
We also compared against state-of-the-art end-to-end methods for automatic image enhancement.
Results show that K-nearest neighbors is an effective strategy for predicting color grading decisions, outperforming state-of-the-art end-to-end methods for image enhancement and demonstrating that modeling compact core parameters is more tractable and interpretable than end-to-end learning.
These core parameters may generalize across the broader landscape of adaptive image processing (e.g., for the development of new automatic exposure, tone mapping, and style transfer methodologies) as these tasks can benefit from perceptually meaningful segmentation of the intensity distribution.
The methodology proposed here also is also promising to be applied to local image regions and motion pictures.


\bibliographystyle{plain}
\bibliography{template}

\end{document}